\documentstyle[prl,aps,epsf]{revtex}
\draft
\begin{document}
\twocolumn[\hsize\textwidth\columnwidth\hsize\csname@twocolumnfalse\endcsname

\title{Contact tracing and epidemics control in social
networks}

\author{Ramon Huerta$^{1,2}$, Lev S. Tsimring$^{1}$}
\address{(1) Institute for Nonlinear Science,
University of California, San Diego, La Jolla, CA  92093-0402}

\address{(2) GNB,
E.T.S. de Ingenier\'{\i}a Inform\'{a}tica,
Universidad Aut\'{o}noma de Madrid, 28049 Madrid (SPAIN).}

\date{\today}

\maketitle

\begin{abstract}
A generalization of the standard susceptible-infectious-removed (SIR) 
stochastic model for epidemics in sparse random networks is
introduced which incorporates contact tracing in addition to random
screening. We propose a deterministic mean-field description which 
yields quantitative agreement with stochastic simulations on random graphs.  
We also analyze the role of contact tracing in epidemics control in 
small-world networks and show that its effectiveness grows as the
rewiring probability is reduced.
\end{abstract}

\pacs{PACS numbers: 87.23.Ge, 87.10.+e}

\narrowtext
\vskip1pc] 

Properties of complex networks recently attracted much attention in
physical community \cite{networks}. Although perhaps it was
prompted by the advent of the Internet and World-Wide Web, the
importance of this subject goes far beyond computer networks. Indeed,
daily commute, power and goods traffic, wired and wireless
communication, disease spreading occur within certain physical or social
networks. The theory of disease spreading, which is known as
mathematical epidemiology, has a long and rich history (see, e.g.
\cite{bailey}).  However, until recently the epidemiological studies have been
mostly concerned with so-called mean-field description of epidemics, in which
it is assumed that 
at any time the probability to get infected is the same for all
individuals. In some other works, spreading of a disease on relatively
simple lattices of individuals has been studied within so called
``forest-fire'' models \cite{johansen}.  Only recently, the studies
which elucidate the role of underlying network structure in the disease
spreading began to appear in the literature
\cite{moore,pastor,rand,andersson}. 

Most of the epidemiological models are based on several simple
assumptions regarding disease contracting and cure. In particular, the
most common mechanism of infection is through a contact with another
infectious individual, and the mechanism of recovery is either
deterministic or purely stochastic with a certain typical time of
recovery. In the simplest Susceptible-Infectious-Susceptible (SIS)
model, a recovered individual immediately becomes susceptible again,
while in a more complicated Susceptible-Infectious-Removed (SIR) model,
cured individuals become immune and effectively excluded from further
dynamics. 

While these models give a good description of evolution of many common
infectious diseases, they usually neglect 
the role of intelligent strategies to stop
nascent epidemics.  Few epidemiological models take into account prevention
strategies such as, for example, mass and ring vaccination
\cite{muller}. In practice, one of the main counter-epidemics
measures is the {\em contact tracing}, when individuals which
have been in contact with an infected (and identified) individual,
are found and thoroughly checked.  It applies, among others, to the
treatment of sexually transmitted diseases, tactics of
law-enforcement organizations trying to uncover criminal or terrorists
networks, cleaning of computer virus infection, etc.  
We only are aware of one
theoretical paper \cite{muller3} where a model of this kind has been
studied. The model \cite{muller3} is based on the assumption that
infection is a slow branching process, while the contract tracing occurs
at a much shorter time scale. This leads to a familiar SIR-type model
with rescaled parameters and similar dynamics. In this Letter we
consider a more realistic model in which infection and contact tracing
occur concurrently, and their interplay determines the 
dynamics of the system. 

{\em Stochastic model}. 
We assume that the population consists of $N$ hosts whose connections to
each other form a fixed graph. The hosts are enumerated with index
$n=1,...,N$. A node $n$ is said to have a degree $k(n)$ if it is
connected to $k$ other hosts. In case of random graphs the degree
distribution is Poissonian with a certain mean degree $K=\langle
k(n)\rangle$.

For simplicity we assume that there in no spontaneous
recovery, an infectious individual can only be disinfected externally
through screening.  Immediately upon disinfecting, the individual
becomes {\em traced} (T) for a certain period of time during which its
neighbors are checked for possible infection.  After that time, the
individual spontaneously becomes removed, and its neighbors are no
longer traced.  

{\bf Infection} $S\rightarrow I$. 
Initially, the whole population except for one host is assumed to be
susceptible to infection. The probability of host
infection depends on the state of its nearest neighbors. The
infection dynamics is modeled as a simple contact process: if a
susceptible node $n$ has $k_i(n)$ infectious neighbors, the
probability that it becomes infectious during a small
$\Delta t$ time interval is $\alpha \, k_i(n) \Delta t$.

{\bf Tracing} $I\rightarrow T$.
The process of infection elimination consists in finding infectious hosts
and then curing them. Hosts are being checked with certain probability
$\beta$ which depends on the state of its neighbors.  We postulate that
if an infectious host is checked, it is immediately cured, eliminated or
at least isolated so it cannot infect other hosts.  We introduce two
non-exclusive strategies of checking for infectious hosts: random checking
and contact tracing. Random checking means choosing an arbitrary host
with probability $\beta_r \Delta t$, while contact tracing of host $n$
is done with probability $\beta_{t}k_t(n) \Delta t$ where $k_t(n)$ is
the number of neighbors of $n$ which are in the {\em traced} state $T$.
The random checking process is equivalent to the removal process of
general epidemics\cite{bailey}.

{\bf Removal} $T  \rightarrow R$. With certain probability $\gamma
\Delta t$, traced hosts are transformed into the {\em removed} state, 
in which they also cannot be infected, but they are no longer under 
observation, so they do not initiate contact tracing. 

{\em Stochastic simulations} of the described process were performed
using  an event-driven scheme\cite{bidaux}. It is significantly superior
over synchronous and asynchronous update schemes both in terms of
accuracy and computational speed.  In the event-driven scheme we select
the time lapsed between two consecutive events from a Poisson
distribution with a combined probability of all events (infection,
tracing, recovery, etc.), then choose a node (each node has its own
probability to be chosen depending on its state and the states of its
neighbors), and apply the transition from one state to another according
to the ratio of individual transition probabilities. 

In our simulations with random graph based networks, we typically built
networks with average degree $K=10$ and $1000$ nodes.  For every random
graph we ran $100$ simulations starting every time from a single (but
different in each run) infected host. Then we averaged the results for
$50$ random graphs.  The time bin was $\Delta t=10^{-6}$ for most
simulations.  In all simulations we varied the tracing parameters
$\beta_r$ and $\beta_t$, while the infection constant was set at
$\alpha=0.1$, and the transition rate from $T$ to $R$, $\gamma=0.5$.
The latter parameter is important for the effectiveness of the targeting
elimination, because the longer a node remains in the traced state, the
more probable it is to trace its neighboring infectious nodes. However,
since tracing presumably bears a significant cost, an optimal choice of
the tracing parameters ($\beta_r, \beta_t$, and $\gamma$) for a given
epidemics is an important issue. 

\begin{figure}
\begin{center}
   \epsfxsize=8 cm

   \epsffile{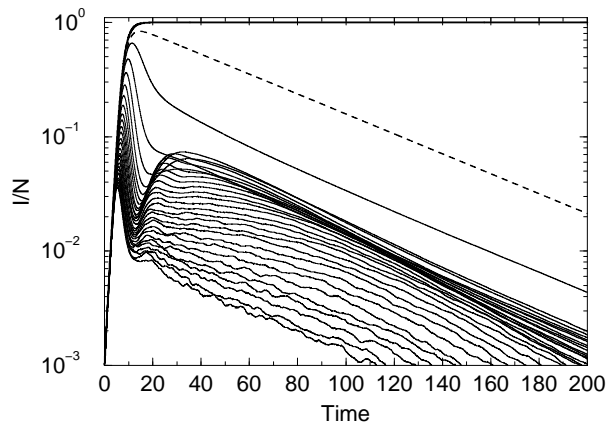}
\end{center}
\caption{Infected population in a random graph of $1000$ nodes and
$K=10$ for $\alpha=0.1, \gamma=0.5$. The solid thick line corresponds to
$\beta_r=0, \beta_t=0$, and the dashed line to $\beta_r=0.02, \beta_t=0$.
Thin lines are for $\beta_r=0.02$ and
$\beta_t=0.1,0.2,\ldots 2.5$. 
\label{fig1} }
\end{figure}

In Fig. \ref{fig1} we present the ``prevalence'' of epidemics (the
fraction of infectious nodes in the whole population $i=I/N$) as a function of
time for several values of $\beta_r$ and $\beta_t$. When $\beta_r=
\beta_t=0.0$ we have a simple $SI$ process, and all the nodes eventually
get infected (thick solid line in Fig. \ref{fig1}). 
Other curves show the fraction of infectious nodes as a function of time
for $\beta_r=0.02$ and different values of $\beta_t$.
The ratio $\alpha/\beta_r$ is chosen to be above the epidemics 
threshold \cite{bailey}.  For $\beta_t=0$ we obtain the classical
$SIR$ process with randomly removed infectives (dashed line in 
Fig. \ref{fig1}). The epidemic eventually saturates, and the
fraction of infectious nodes decays exponentially.
The lower lines display the evolution of the infection fraction for
values of $\beta_t$ ranging from $0$ to $2.5$ with a step value of
$0.1$.  The initial (exponential) phase of the epidemics growth is
nearly independent of $\beta_t$,  because the contact tracing process is
intrinsically nonlinear (it  requires the presence of I-T connected pairs
and therefore only begins after the first infected node is randomly
screened).  As expected, the tracing process significantly reduces 
the magnitude of the epidemics (maximal value of $i$), but at large
times the infection decays with the same exponential rate as for
$\beta_t=0.0$ (again, because we return to the linear regime at small 
$i$). The most interesting feature of the process at large $\beta_t>0.35$
 is the presence of a second maximum of $i$ which indicates a second 
epidemic.  Due to this second epidemic, the percentage of the
infectious population  at large times $t>40$ may actually increase with
increase of $\beta_t$. It means that the range
of $\beta_t$ values from $0.4$ to $0.9$ are not better to control the
epidemics than values smaller than $0.4$. 

{\em Mean-field equations}.
At the first
sight, it seems that the mean-field approach 
cannot be applied to the contact tracing, since it does not
take into account 
the non-uniform distribution of infection in the
population. Nevertheless, a more sophisticated mean-field approach which
operates not only with the mean densities of states, but also with the
densities of links connecting nodes with different states, can be
devised. In the derivation we follow the method of Rand\cite{rand} (see
also \cite{andersson}).  Let us introduce the number of nodes $A$, 
the number of connected pairs $[AB]$ and triples $[ABC]$ of nodes,
where $A, B$ and $C$ stand for any of the types $S,I,T,R$. For example,
the number of connected pairs of infectious and traced nodes is denoted
$[IT]$. Note that $[AB]=[BA]$ and each pair in $[AA]$ is counted
twice. For large $N$, the ratios $A/N, [AB]/N, [ABC]/N$ approach
deterministic limits which we label $a, [ab], [abc]$, respectively. 

The dynamics of the model is described by the following set of rate equations
\begin{eqnarray}
\dot{s}&=&-\alpha [si],\ \ 
\dot{i}=\alpha [si] -\beta_r i - \beta_t [i\tau],
\label{rate_i}\\
\dot{[ss]}&=&-2\alpha [ssi],
\label{rate_ss}\\
\dot{[si]}&=&\alpha ([ssi]-[isi]-[si])-\beta_r [si]-\beta_t [si\tau],
\label{rate_si}\\
\dot{[ii]}&=&2\alpha ([isi]+[si])-2\beta_r [ii]-2\beta_t [ii\tau],
\label{rate_ii}\\
\dot{[i\tau]}&=&\beta_r [ii]+\beta_t([ii\tau]-[\tau i\tau
]-[i\tau])-\gamma[i\tau]-\beta_r[i\tau].
\label{rate_it}
\end{eqnarray}
Here we used the notation $\tau=[T]/N$ to avoid confusion between the
density of traced nodes and time $t$. Note that we omit here the
equations for $\tau, [\tau\tau], [s\tau]$, as well as any combinations
involving the removed state, since they do not affect the dynamics of
the infectious population.

The meaning of these equations is rather straightforward. For example,
the terms in the r.h.s. of the last equation can be explained as
follows.  A $(p,q)$ pair becomes $[i\tau]$ through random screening of
the infectious node $q$ in an $[ii]$ pair, or through contact tracing of
node $q$ from node $r$ in a $(p,q,r)$ triple in the state $[ii\tau]$. On
the other hand, we can lose an $[i\tau]$ pair by contact tracing of the
$p$ node in a $[\tau i\tau]$ triple $(r,p,q)$, by direct tracing $p$ by
$q$, by removing of $q$, or by random screening of $p$. Other equations
can be obtained from similar arguments. 

This set of equation is not closed, as the equations for the
pair densities contain triple densities. We need to introduce a
closure rule. Similarly to Refs.\cite{rand,andersson}, we can
use the approximation $[abc]=[ab][bc]/b$,
which follows from the condition that the influence of a node on the
state of its second neighbor in a triple is negligible\cite{condition}.

Using this closure rule, we arrive at the following set of
equations
\begin{eqnarray}
\dot{s}&=&-\alpha s \hat{i},
\label{rate_s1}\\
\dot{i}&=&\alpha s\hat{i} -\beta_r i - \beta_t i\hat{\tau},
\label{rate_i1}\\
\dot{\hat{i}}&=&(\alpha Ks-\alpha-\beta_r)\hat{i}-\beta_t\hat{i}\hat{\tau},
\label{rate_si1}\\
\dot{\hat{\tau}}&=&(\beta_t\hat{p}-\beta_t-\gamma-\alpha\frac{s}{i}\hat{i})
\hat{\tau}+\beta_r\hat{p},
\label{rate_ii1}\\
\dot{\hat{p}}&=&\alpha\frac{s}{i}(2\hat{i}+2-\hat{p})\hat{i}-
(\beta_r+\beta_t\hat{\tau})\hat{p}.
\label{rate_it1}
\end{eqnarray}
where $\hat{i}=[is]/s$ is the mean number of infectious neighbors per
susceptible node, $\hat{\tau}=[i\tau]/i$ is the mean number of traced
neighbors per infectious node, and $\hat{p}=[ii]/i$ is the mean number of
infectious neighbors of an infectious node. Notice that the equation for
$[ss]$ dropped out as $[ss]=Ks^2$ at all times. We used the initial conditions
$s(0)=1-i_0, i(0)=i_0, \hat{i}(0)=(K-1)i_0,
\hat{\tau}(0)=0, \hat{p}(0)=0$ which correspond to a small set of
disconnected infectious nodes. 

\begin{figure}
\begin{center}
   \epsfxsize=7 cm
   \epsffile{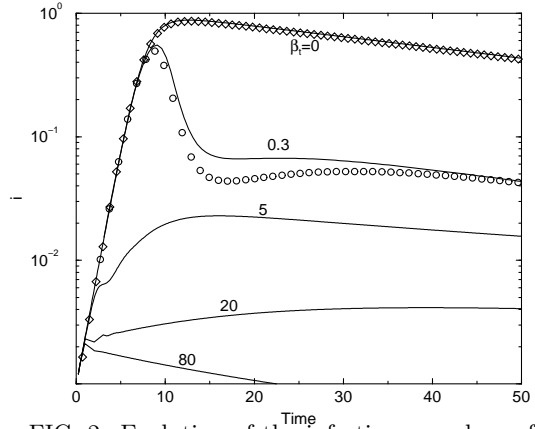}
\end{center}
\caption{Evolution of the infection prevalence for different $\beta_t$, mean-field model (lines) and
stochastic simulations (symbols),  $\alpha=0.1, K=10,
\beta_r=0.02, \gamma=0.5$.
\label{sim} }
\end{figure}

During the early stage of an epidemic the
contact tracing can be neglected ($\tau=0$), and Eqs.
(\ref{rate_s1})-(\ref{rate_it1}) are reduced to a set of three equations for
$s,\ i,\ \hat{i}$
which coincide with the model has been studied in \cite{rand,andersson}.
Independently of $\beta_t$, the initial epidemics growth is
characterized by the basic reproduction number
$K\alpha/(\alpha+\beta_r)$.  However, as the number of traced
individuals grows, the growth rate is reduced and the epidemics is
saturated. In Fig.\ref{sim} the dynamics of the epidemics calculated
from Eqs.(\ref{rate_s1})-(\ref{rate_it1}) are shown for different 
values of $\beta_t$.
As seen from the figure, the maximum number of infectious nodes is
drastically reduced with increase of $\beta_t$. In the same figure we
show the results of direct stochastic simulations for $\beta_t=0,\
0.5$\cite{note}.
The most important
question is whether the contact tracing is capable of arresting the
exponential growth of the epidemics before it engulfs a finite portion
of the total population. 

\begin{figure}
\begin{center}
   \epsfxsize=7 cm
   \epsffile{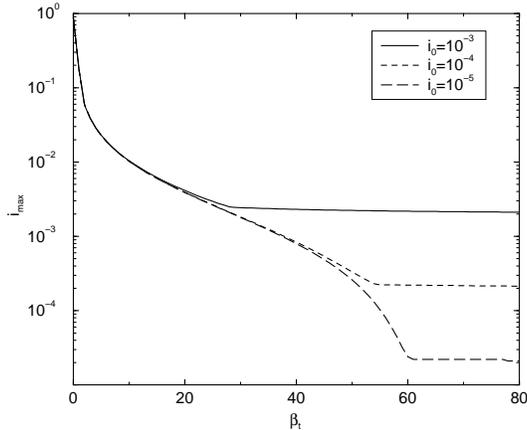}
\end{center}
\caption{Maximum number of infectious nodes vs. $\beta_t$ for different
$i_0$.
Parameters are the same as in Fig.\protect\ref{sim}
\label{imax} }
\end{figure}

To answer this question, we consider the limit of small epidemics in a
large populations ($i,\hat{i}<<1$), then we can set $s(t)=1$ and drop
Eq.(\ref{rate_i1}). We also observe that $\hat{i}/i=K-1$ and drop
Eq.(\ref{rate_i1}) as well.  A simple calculation shows that  the
critical value of $\beta_t$ at which the
exponential growth of infection is arrested,
\begin{equation}
\beta_{cr}=\frac{(\alpha(K-1)+\gamma)(\alpha(K-1)-\beta_r)}{\beta_r}.
\label{beta_cr}
\end{equation}
For $\beta>\beta_{cr}$, epidemic remains small at all times, and so 
there is no major outbreak of the epidemic, 
the maximum number of infectious nodes is independent of the population size, and
depends on the initial size of infection $i_0$.
For the parameter values of our stochastic simulations,
$\alpha=0.1,K=10,\beta_r=0.02,\gamma=0.5$, we obtain $\beta_{cr}=61.6$. 
In Fig.\ref{imax} we show the maximum number of infectious nodes during
the epidemic as a function of $\beta_t$ at different initial epidemic
sizes $i_0$. 
In agreement with the above argument, for $\beta_t<\beta_{cr}$ the value
of $i_{max}$ is almost independent on $i_0$ until $i_{max}\approx
2i_{0}$ at a certain $\beta_t^*$,
after which it remains nearly constant. The value of $\beta_t^*$
approaches $\beta_{cr}$ for $i_0\to 0$.

\begin{figure}
\begin{center}
   \epsfxsize=8 cm
   \epsffile{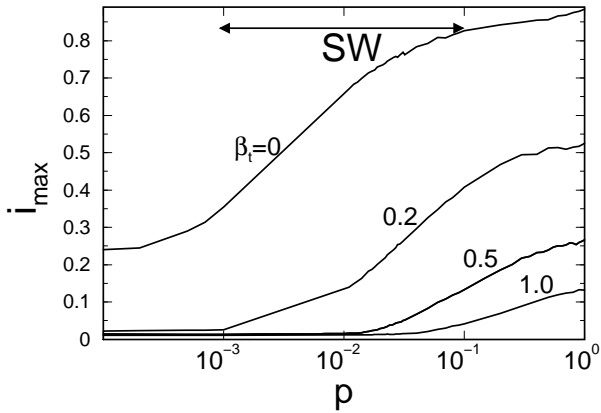}
\end{center}
\caption{The maximum prevalence as a function of the rewiring
probability $p$. Effectiveness of the contact
tracing becomes very significant in the small-world regime.
\label{figsw2} }
\end{figure}

It is easy to see that the epidemic threshold depends only on the
average node degree and  not on the specific topology of the underlying
graph. However, the subsequent development of the epidemics should be
significantly affected by the network structure, in particular by the
average minimum path length and the clustering coefficient\cite{networks}.  
Low average minimum path
means that any node in the graph can be exposed to infection in a
short time after an epidemics begins.  If the clustering coefficient is
large, the infection propagates faster within a certain
community but then it make it easier to trace the
epidemics.  Sparse
random networks studied  above represent a particular class of
networks with short average minimal path and small clustering
coefficient.  Many social networks are
characterized by a relatively large clustering coefficient while keeping
the average minimal path low.  We studied numerically the effect of the
network structure on the contact tracing of epidemics within the
small-world model \cite{watts}. Changing
the re-wiring probability $p$ allows us to scan the range of networks
from regular ($p=0$) to random ($p\to 1$) through the small-world range
($0.001<p<0.1$) which exhibits a short average minimal path and a large
clustering coefficient typical for many social networks.  We used the
same number of nodes and  edges as for the random graph simulations, and
fixed the parameter values at $\alpha=0.1, \beta_r=0.02,$ and
$\gamma=0.5$.  Fig.\ref{figsw2} shows the dependence of the maximum
prevalence $i_{max}$ on $p$ for several different $\beta_t$.  As we can
see, $i_{max}$ changes mostly within the SW range ($0.001<p<0.1$) where
the clustering coefficient and the average path undergo large
variations.

In this Letter we studied the role of contact tracing as a part of the
epidemics control strategy in complex networks. We demonstrated that by
applying this strategy, a major outbreak can be significantly reduced or
even eliminated at a small additional cost.  Based on the pair
correlation approach due to Rand \cite{rand}, we developed the
mean-field model of contact tracing for the case of random graphs.
We also studied the influence of network topology on the contact tracing
using the small-world model with variable re-wiring probability $p$, and
found that its effectiveness grows as the reviring probability is
reduced. The main change occurs within the small-world regime at 
$p\sim 10^{-2}$.

This work was supported by the U.S. DOE grant
DE-FG03-96ER14592, ARO MURI-ARO grant DAAG55-98-0269, MCyT BFI2000-157,
and NATO Collaborative Linkage Grant PST.CLG.978512.

\vspace{-0.3in}
\references
\bibitem{networks} 
M. E. J. Newman, J. Stat. Phys. 101 (3-4): 819 (2000); 
S. H. Strogatz, Nature, {\bf 410 (6825)}: 268 (2001);
R. Albert, A. Barabasi, cond-mat/0106096 (2001).
\bibitem{bailey}N.T.J.Bailey, {\em The mathematical theory of infectious
diseases}, Charles Griffin \& Co., London, 2nd ed., 1975.
\bibitem{johansen} A. Johansen, Physica D, {\bf 78}, 186 (1994).
\bibitem{moore}C.Moore and M.E.J.Newman, Phys. Rev. E, {\bf 61}, 5678
(2000).
\bibitem{pastor}R.Pastor-Satorras and A.Vespignani, Phys. Rev. Lett.,
{\bf 86}, 3200 (2001).
\bibitem{rand}D.Rand, in: Advanced Ecological Theory. Principles and
Applications (ed. J. McGlade), pp. 100-142, Blackwell Science, 1999. 
\bibitem{andersson} H.Andersson and T.Britton,  Epidemics: Stochastic
models and their statistical analysis, Springer Lectire Notes in
Statistics, {\bf 151}, Springer-Verlag, Ney York (2000).
\bibitem{muller} J. M\"uller, SIAM J. Appl. Math. {\bf 59(1)}, 222
(1998); .J. Mathermatical Biology {\bf 41}, 143-171 (2000).
\bibitem{muller3} J. M\"uller, M. Kretzschmar, K. Dietz, Mathematical
Biosciences {\bf 164} 39 (2000).
\bibitem{bidaux}R.Bidaux, B.D.Lubachevsky, and V.Privman, Phys. Rev. A, 
{\bf 44}, 6213 (1991).
\bibitem{condition} This
condition becomes asymptotically correct for $N\to\infty$, as the
probability of a triple to form a connected triangle is $O(N^{-1})$. However,
for other types of networks, such as lattices or small-world networks,
this condition clearly is not satisfied.
\bibitem{note} For larger values of $\beta_t$ the number of infectious and
traced nodes become small, and the agreement between the theory and 
simulations becomes only qualitative. 
\bibitem{watts} D. J. Watts, S. H. Strogatz, Nature, {\bf 393} 440
(1998).
\end{document}